\documentclass[pra, twocolumn, amsmath,amssymb, superscriptaddress, nofootinbib, floatfix ]{revtex4-1}

\usepackage{multirow}
\usepackage{graphicx}
\usepackage{amsfonts,tikz}  
\usetikzlibrary{matrix}
\usepackage{epstopdf}
\usepackage[version=3]{mhchem}
\usepackage{color}
\usepackage{hyperref}
\usepackage{enumitem}
\usepackage[scr=boondoxo,scrscaled=1.25]{mathalfa}
\usepackage{scalerel}
\usepackage{esint}

\usepackage[euler]{textgreek}

\hypersetup{%
  pdfpagemode=None, 
  pdfstartpage=1,
  pdfstartview=FitH,
  pdfmenubar=true,
  pdftoolbar=true,
  colorlinks = true,
  linkcolor=blue,
  citecolor=blue,
  urlcolor = blue,
  bookmarksopen=false
}

\newcommand{\Bra}[1]{\ensuremath{\left\langle #1 \right\vert}}
\newcommand{\KetBra}[2]{\Ket{#1}\kern-0.1em\Bra{#2}}
\newcommand{\Ket}[1]{\ensuremath{\left\vert #1 \right\rangle}}

\definecolor{Pantone268}{cmyk}{0.82,1.0,0.0,0.12}
\definecolor{HunterOrange}{cmyk}{0.0,0.55,1.0,0.0}

\begin{document}

\title{Spectral properties of high-order harmonic radiation enhanced by XUV-driven electron-hole dynamics}

\date{\today}

\author{R.~Esteban Goetz}
\email{egoetz@uconn.edu}
\affiliation{Department of Physics, University of Connecticut, 196A Auditorium Road, Unit 3046, Storrs, CT 06269}
\author{Anh-Thu Le}
\affiliation{Department of Physics, University of Connecticut, 196A Auditorium Road, Unit 3046, Storrs, CT 06269}

\begin{abstract}

We analyze the spectral properties of high-order harmonic radiation with photon energies
extending beyond the regular cutoff energy in standard high-order harmonic generation. The extension of the regular harmonic cutoff results from infrared (IR)-driven recombination of valence photoelectrons into a cationic core hole created by extreme-ultraviolet (XUV) excitation of inner-shell electrons into the transient valence hole in a combined XUV+IR configuration [Buth \emph{et al.}, Opt. Lett. \textbf{36}, 3530 (2011)].
We show that the microscopic dipole phase at the extended harmonic frequencies is sensitive to the relative IR-XUV delay and IR intensity, whereas the corresponding signal intensity drops significantly for chirped XUV pulses with poor temporal coherence. We discuss the impact of such sensitivity on the macroscopic harmonic radiation, whereby decoherence among the dipole emitters may lead to further signal suppression.
\end{abstract}

\maketitle

\section{Introduction} 

High-order Harmonic Generation (HHG) is a highly nonlinear photoconversion process by which intense laser light is used to generate high-energy electromagnetic radiation
comprising high-order harmonics of the incident laser frequency~\cite{corkum2007attosecond,Agostini2004,salieres1995}. HHG is routinely used to generate phase-locked attosecond pulsed trains~\cite{paul2001observation,corkum2007attosecond} spanning the extreme ultraviolet and
the soft x-ray spectrum~\cite{paul2001observation,hentschel2001}. 

The 
three-step model~\cite{kulander1993,Lewestein1994}
provides a semiclasical description of the harmonic emission process at the microscopic level: when atoms or molecules are exposed to an intense laser field, 
the photoionized electrons are accelerated before recolliding with their parent ions.
The energy
gained by the photoelectrons in the laser field during the excursion time~\cite{IvanovPRL2013} is then released in the form of high-energy photons~\cite{corkum2007attosecond,Agostini2004}. The resulting radiation spectrum is characterized by a plateau of harmonics followed by abrupt decay of the intensity beyond cutoff energy~\cite{Lewestein1994,Corkum1994}.

Several mechanisms yielding to an extension of the energy cutoff have been proposed~\cite{CarreraPRA2007,Fu2023,BandraukPRA2002,ShanPRA2001,Buth2011,Buth2013,Buth2015}. In particular, extension mediated by radiative recombination of valence photoelectrons into a core hole created by XUV resonant core-valence shell excitation~\cite{Buth2011,Buth2015}. A formalism describing this process was derived based on a two-active electron approximation with no Coulomb interaction between the electrons and no electron correlation in the ground and final states~\cite{Buth2011,Buth2013,Buth2015} and applied to the $3d\rightarrow 4p$ resonance transition in krypton~\cite{Buth2011}. Extension of the
cutoff then results from radiative recombination of a $4p$ photoelectron into the $3d$ hole created by the resonant XUV pulse which refills the transient $4p$ hole~\cite{Buth2011}.

 Despite the dramatic shift of the cutoff predicted by the theory for the case of XUV-assisted HHG in krypton~\cite{Buth2011,Buth2013,Buth2015}, the solely experimental realization reported to date~\cite{Tross2022} and  performed at FLASH~\cite{FLASH2010} could not observe 
a clear and unambiguous uplift of the harmonic cutoff, even though the theory 
considered limited temporal coherence of SASE pulses~\cite{Buth2011}.  

In this work, we analyze the spectral properties of the harmonics in the extended spectral region in an attempt to explain the discrepancy between the prediction of the theory~\cite{Buth2011,Buth2013,Buth2015} and experimental observations~\cite{Tross2022}. To this end, we extend the theoretical considerations of Refs.~\cite{Buth2011,Buth2015} by
incorporating multielectron  and 
interchannel effects in the context of the Time-Dependent Configuration-Interaction Singles (TDCIS) formalism~\cite{RohringerPRA2006,GreenmanPRA2010},
which is particularly
suitable for investigating the electron-hole dynamics resulting
in the cut-off extension investigated in this paper. In particular, we analyze the sensibility of the atomic dipoles $\hat{d}(\omega_q)$ in the extended spectral region, on e.g., the XUV and IR relative delay and XUV chirp in the 
interest of estimating the impact of lost of relative temporal coherence between the propagating XUV and IR pulses on the macroscopic yield. In fact, different refractive indices for the IR and XUV, $n_{\mathrm{IR}}$ resp. $n_{\mathrm{XUV}}$, results in a position-dependent delay 
$\tau=(n_{\mathrm{XUV}}-n_{\mathrm{IR}})\Delta z/c$ between them
that depends longitudinally on the propagation distance $\Delta z$. We show that the phase of the microscopic dipoles $\hat{d}(z_k,\omega_q)$ located along the propagation axis $z_k$ is a linear function of the XUV-IR delay $\tau_k=(n_{\mathrm{IR}}-n_{\mathrm{XUV}})(z_k-z_0)/c$ 
with the slope (sensitivity) determined by the 
core-valence shell energy difference, $\Delta E$. Using Argon as a prototype, we analyze the impact of the difference in the propagation speed of the XUV and IR pulses due to their different refractive indices on the macroscopic HHG signal in the extended plateau region as well as the impact of XUV absorption on the 
macroscopic signal. Finally,  we investigate the impact of  partially incoherent XUV pulses on the extended spectrum. We find a suppression that depends on the statistically-averaged coherence time $\langle\tau_c\rangle$, with a five-fold suppression for
$\langle\tau_c\rangle=1\,$fs.  To keep calculations tractable, we perform proof-of-principle calculations  for the $3s\rightarrow 3p$ transition in  argon (8 active electrons) but also show results for the $3d\rightarrow 4p$ transition in krypton with $18$ active electrons.

\section{Theoretical model} 
\label{sec:theory}

\subsection{Many-electron dynamics}
\label{subsec:theory_tdcis}

Our method for treating the multi-electron dynamics in atoms
is based on the time-dependent configuration-interaction singles (TDCIS) formalism as 
described in Ref.~\cite{GreenmanPRA2010} 
In the configuration-interaction singles basis, many-electron state vector is expressed according to,

\begin{eqnarray}
\label{eqn:linear_comb0}
    |\Psi(t;\boldsymbol{r}_n)\rangle = \Tilde{C}_0(t;\boldsymbol{r}_n)|\Phi_0\rangle + \sum_{i,a}\Tilde{C}^a_{i}(t;\boldsymbol{r}_n)|\Phi^a_i\rangle\,.
\end{eqnarray}
Here, $\boldsymbol{r}_n$ specifies the atom location and defines the position at which the IR and XUV driving fields are evaluated in the sample~\cite{Kulander1990}. $|\Phi_0\rangle  =  \prod^{N/2}_{i=1}\hat{c}^{\dagger}_{i,-}\hat{c}^{\dagger}_{i,+}|0\rangle$ represents the Hatree-Fock ground state and  $|\Phi^a_i\rangle  =\sum_{\sigma=\pm }\hat{c}^{\dagger}_{a,\sigma}\hat{c}_{i,\sigma} |\Phi_0\rangle/\sqrt{2}$
the singly particle-hole excitations
with $|0\rangle$ the vacuum state. The operator $\hat{c}^{\dagger}_{i,\sigma}$ ($\hat{c}_{i,\sigma}$) creates (anihilates) electrons in the spin-orbital $|\varphi_{i,\sigma}\rangle$, labeled by the index $i$ and spin $\sigma$, i.e, $\hat{c}^{\dagger}_{i,\sigma}|0\rangle = |\varphi_{i,\sigma}\rangle$, with $|\varphi_{i,\sigma}\rangle$ the eigenstates of the Fock operator. 
\medskip

The time-dependent Schr\"odinger equation reads,
\begin{eqnarray}
\label{eqn:com}
i \dfrac{\partial}{\partial t}|\Psi(t;\boldsymbol{r}_n)\rangle = 
\left( \hat{H}_0 + \hat{H}_1+ \hat{H}_{L}(t;\boldsymbol{r}_n) \right) |\Psi(t,\boldsymbol{r}_n)\rangle\,,
\end{eqnarray}
where $\hat{H}_0\!=\!\sum_j  \big[-\boldsymbol{\nabla}^{2}_j/2 \!-\!Z/|\boldsymbol{r}_j \!-\! \boldsymbol{r}_n| + \hat{V}_{MF}(\boldsymbol{r}_j) \big]$ contains all one-particle operators including the  mean-field potential $\hat{V}_{MF}$~\cite{RohringerPRA2006}. $\hat{H}_1 \!=\!\sum_{j>i} 1/|\boldsymbol{r}_i\!- \!\boldsymbol{r}_j| \!- \!\hat{V}_{MF}$ accounts for all two-particle electron-electron interactions beyond $\hat{V}_{MF}$~\cite{PabsPRL2013,RohringerPRA2006}. $\hat{H}_1$  gives rise to interchannel couplings between the photoelectron and cation by mixing different ionization channels, i.e., hole states~\cite{PabsPRL2013}. Finally, $H_{L}(t;\boldsymbol{r}_n)=-\hat{\boldsymbol{d}}\cdot\boldsymbol{E}(t;\boldsymbol{r}_n)$ with $\hat{\boldsymbol{d}}$ the dipole operator, and
\begin{subequations}
\begin{eqnarray}
\label{eqn:fields}
    \boldsymbol{E}(\boldsymbol{r}_n,t) = \boldsymbol{E}_{\mathrm{IR}}(\boldsymbol{r}_n,t) + \boldsymbol{E}_{\mathrm{XUV}}(\boldsymbol{r}_n,t-\tau)\,,
\end{eqnarray}
\end{subequations}
the driving fields,
where $\boldsymbol{E}_{\mathrm{IR}}(\boldsymbol{r}_n,t)$ and $\boldsymbol{E}_{\mathrm{XUV}}(\boldsymbol{r}_n,t)$ denote the IR and XUV  components with central frequencies $\omega_0$ and $\omega_{\mathrm{XUV}}$, respectively, and $\tau$ the XUV-IR time delay. 
The time-dependent coefficients in Eq.~\eqref{eqn:linear_comb0} are obtained by integrating the coupled equations of motion~\cite{GreenmanPRA2010}
\begin{subequations}
\label{eqn:eqn_motion}
\begin{eqnarray}
    \dot{C}_0(t;\boldsymbol{r}_n) &\!=\!& 
     i E(t;\boldsymbol{r}_n)\sum_{i,a} \langle\Phi_0|\hat{\boldsymbol{d}}\cdot\hat{\boldsymbol{\mathrm{e}}}|\Phi^a_i\rangle\,C^a_i(t;\boldsymbol{r}_n)\,,
     \end{eqnarray}
     for the Hartree-Fock gound state, and
     \begin{eqnarray}
    \dot{C}^a_i(t;\boldsymbol{r}_n) &\!=\!& i E(t;\boldsymbol{r}_n)\, \langle\Phi^a_i|\hat{\boldsymbol{d}}\cdot\hat{\boldsymbol{\mathrm{e}}}|\Phi_0\rangle\, 
     C_0(t;\boldsymbol{r}_n)\\[0.3cm]
    &+& i E(t;\boldsymbol{r}_n)\sum_{j,b} \langle\Phi^a_i|\hat{\boldsymbol{d}}\cdot\hat{\boldsymbol{\mathrm{e}}}|\Phi^b_j\rangle\, C^a_i(t;\boldsymbol{r}_n)\nonumber\\[0.15cm]
      &+& i E(t,\boldsymbol{r}_n)\sum_{j,b} \langle\Phi^a_i|\hat{H}_1|\Phi^b_j\rangle\,C^b_j(t;\boldsymbol{r}_n)\,,\nonumber
\end{eqnarray}
\end{subequations}
for the singly particle-hole excitations,
with $\hat{\boldsymbol{\mathrm{e}}}$ the unit polarization vector (same for IR and XUV), and
where we have defined $\tilde{C}_0(t;\boldsymbol{r}_n)\!=\!\exp[-i\varepsilon_0\, t]\, C_0(t;\boldsymbol{r}_n)$ and $\tilde{C}^a_i(t)\!=\! \exp[-i(\varepsilon_a - \varepsilon_i + \varepsilon_0)\, t]\,C^a_i(t;\boldsymbol{r}_n)$ with $\varepsilon_0$ the Hartree-Fock energy 
and $\varepsilon_{i/a}$ the orbital energy with $\hat{H}_0|\varphi_{{i/a},\sigma}\rangle = \varepsilon_{i/a}|\varphi_{{i/a},\sigma}\rangle$. For coherent XUV and IR fields, we consider a plane-wave description of the incident fields propagating in the $z$ direction,
\begin{subequations}
    \begin{eqnarray}
\label{eqn:plw}
E_{\mathrm{XUV}}(\boldsymbol{r},t)\!=\! G_{1}(z-v_{1} t)\cos\big[\omega_{\mathrm{XUV}} (t-z/v_{1}) + \varphi_{1}\big]\!,\,\,\,\,
\end{eqnarray}
for the incident XUV field, and
\begin{eqnarray}
\label{eqn:plww}
E_{\mathrm{IR}}(\boldsymbol{r},t)\!=\! G_{0}(z-v_{1} t)\,\cos\big[\omega_{\mathrm{0}} (t-z/v_{0}) + \varphi_{0}\big] \,,
\end{eqnarray}
\end{subequations}
for the IR field,
with  $\varphi_q$ the carrier-envelope phase (CEP), $v_q=c/n_q$ the group velocity with $n_q$ the refractive index;
 $G_q(z-v_q t) = \exp [-(z-v_q t )^2/(c^2\tau^2_q)]$ a Gaussian envelope with $\tau_q=\sigma_q/2\log(2)$ where $\sigma_q$ denotes the temporal full width at half maximum (FWHM) of the intensity profile and $q=0,1$. Unless otherwise specified, we let $n_1\!=\!n_0\!=\!1$ and $\boldsymbol{r}_n =(x_0,y_0,z_0)= (0,0,0)$.  
 
 To evaluate the harmonic radiation we consider the dipole acceleration given by the Ehrenfest theorem,
\begin{eqnarray}
\label{eq:def_dipa}
    \ddot{\boldsymbol{d}}(t;\boldsymbol{r}_n)=-\langle\Psi(t;\boldsymbol{r}_n)|\boldsymbol{\nabla} H(\boldsymbol{r}-\boldsymbol{r}_n,t)|\Psi(t;\boldsymbol{r}_n)\rangle
\end{eqnarray}
 with $\boldsymbol{\nabla}$ denoting the gradient operator and 
 $H(\boldsymbol{r}-\boldsymbol{r}_n,t)$ the full Hamiltonian appearing in Eq.~\eqref{eqn:com}.

\subsection{Partially coherent XUV pulses}
\label{subsec:partially_coherent}
To model partially coherent XUV pulses we follow the Partial-Coherence Method (PCM) of Ref.~\cite{Pfeifer2010}. At the sample entrance point $\boldsymbol{r}_0=(x_0,y_0,z_0)$, we describe the partially-coherent incident XUV pulse according to,
\begin{eqnarray}
    \label{eqn:pcm}
   E^{(PC)}_{\mathrm{XUV}}(\boldsymbol{r}_0,t)&=& G_1(z_0-v_1 t)\,
    \int^{\infty}_{0}\!\! d\omega\, \hat{h}(\omega-\omega_{XUV})\nonumber\\[0.1cm]
    &&\times\,\cos\left[\omega_1 \left(t-z_0/v_1\right) + \Tilde{\varphi}_{PC}(\omega)\right]\!,\quad\quad
\end{eqnarray}
where $\hat{h}(\omega-\omega_{\mathrm{XUV}})$ is a Gaussian function of the (circular) frequency $\omega$ centered at the XUV mean photon energy $\omega_{\mathrm{XUV}}$ and of  amplitude $2\sqrt{I_0(\omega)}$, with
$I_0(\omega)$ the spectral density; $\Tilde{\varphi}_{PC}(\omega)$ is a discrete random spectral phase function generated by a set of independent random numbers with values between $\pi$ and $\pi$, and $G_1(z_0-v_1 t)$ the Gaussian function defined in Eq.~\eqref{eqn:plw}.

The
the random function $\Tilde{\varphi}_{PC}(\omega)$
is not a function of position.  As a result, the translation property
$E^{(PC)}_{\mathrm{XUV}}(\boldsymbol{r},t) =E^{(PC)}_{\mathrm{XUV}}(\boldsymbol{r}_0,t-(z-z_0)n_1/c)$ holds
as it also holds for the coherent $E_{\mathrm{XUV}}$ pulse in Eq.~\eqref{eqn:plw}
despite the randomness of the function $\Tilde{\varphi}_{PC}(\omega)$. 
The coherence time $\tau_c$ associated to coherent or partially-coherent fields are evaluated in accordance to the definition~\cite{Saldin2008,Mandel1995}
\begin{eqnarray}
\label{eqn:tau_c}
    \tau_c = \int |g_1(\tau)|^2\,d\tau\,,
\end{eqnarray}
where $g_1(\tau)$ is the first-order time-correlation function for a stationary random process which can be written as $g_1(\tau)=\int e^{i\omega\tau}\, |E(\omega)|^2\, d\omega$.

\begin{figure}[!t]                                                                      
\centering                                                                       
\includegraphics[width=0.99\linewidth]{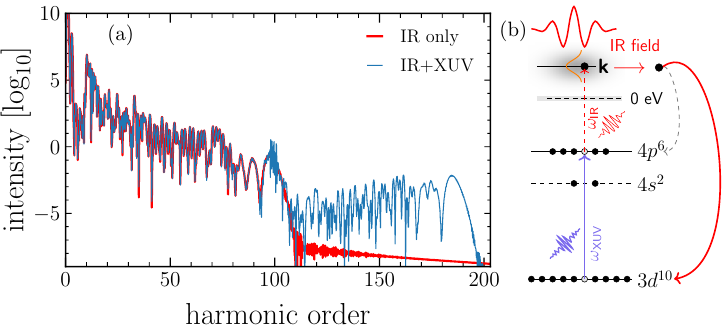}                    
\caption{(a) Dipole spectrum in krypton in the presence (blue lines) and absence (red lines) of the resonant XUV field. The XUV pulse has a duration of $1$ fs (FWHM), peak intensity of
$2\times 10^{12}\mathrm{Wcm}^{-2}$ and central frequency $\omega_{\mathrm{XUV}}=89.78$ eV, corresponding to Hartree-Fock $4p$ and $3d$ orbital energy difference in neutral krypton. The IR field has a duration of $8$ fs, central frequency of $1.0332$ eV (1.2\,$\mu m$) and peak intensity of $2\times 10^{14}\mathrm{Wcm}^{-2}$. (b) Schematic of the XUV-driven charge dynamics resulting in the emission of high-order harmonics lying beyond the regular harmonic cutoff region, see text.}        
\label{fig:figure_krypton} 
\end{figure} 

\section{Numerical results}

\subsection{Coherent XUV pulses}
\label{sec:coherent}

\subsubsection{Extended spectrum in Krypton}

We first apply the TDCIS formalism to krypton to confirm the harmonic cutoff extension 
predicted in Ref.~\cite{Buth2011,Buth2015}. 
TDCIS calculations were performed using the XCID program package~\cite{GreenmanPRA2010} with 18 active electrons (ten electrons in the $3d$ shell, two in $4s$  and six in the $4p$ valence orbital). Hartree-Fock eigenstates were obtained with a pseudospectral radial grid of $220$~bohr with 800 radial points and Moebius mapping parameter $\zeta=0.5$ to optimize the density of points near the origin, a maximum angular momentum $L_{max}=100$, and a complex absorbing potential (CAP) of strength $\eta=0.01$ starting at $200$~bohr to avoid reflections off the grid boundaries. Integration of Eq.~\eqref{eqn:eqn_motion} were performed with the short-iterative Lanczos algorithm with a time step of $0.02$~a.u. Field parameters are given in caption of Fig.~\ref{fig:figure_krypton}.
\begin{figure}[!t]                                                                      
\centering                                                                       
\includegraphics[width=0.99\linewidth]{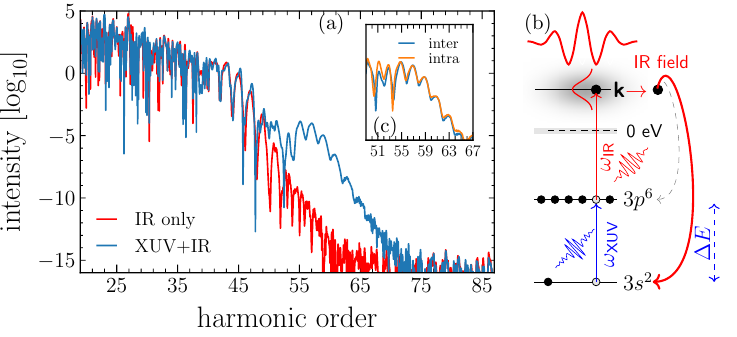}                    
\caption{(a) Dipole spectrum in argon in the presence (blue lines) and absence (red lines) of the resonant XUV field. The XUV pulse has a duration of $2$ fs (FWHM), peak intensity of
$2\times 10^{12}\mathrm{Wcm}^{-2}$ and central frequency $\omega_{\mathrm{XUV}}=18.67$ eV, corresponding to Hartree-Fock $3p$ and $3s$ orbital energy difference in neutral argon. The IR field has a duration of $22$ fs, central frequency of $1.44$ eV (861 nm) and peak intensity of $2\times 10^{14}\mathrm{Wcm}^{-2}$. (b) Schematic of the hole dynamics resulting in the extended harmonic region, see text.}               
\label{fig:figure1} 
\end{figure} 

Figure~\ref{fig:figure_krypton}(a) shows the harmonic yield, hereafter defined as the squared norm of the Fourier transform of Eq.~\eqref{eq:def_dipa}, in the absence (red) and presence (blue) of the resonant XUV field. In the IR-only case, the cutoff energy
$\omega^\prime_c=|\varepsilon_{4p}| + 3.17 E^2_{0}/4\omega^2_0=99.38\,$eV in Fig.~\ref{fig:figure_krypton}(a). Switching on the XUV (zero delay) results in a dramatic extension of the cutoff (blue), from $\omega^\prime_c=99.38\,$eV to $\omega_c= 189.2\,$ eV.  The extension $\omega_c-\omega^\prime_c=89.82\,$eV
corresponds to the $4p-3d$ Hartree-Fock orbital energy difference. Keeping the XUV+IR configuration but freezing the  $3d$ orbital  ($4s$ and $4p$ active orbitals) results in the energy cutoff $\omega^\prime_c$. The TDCIS calculations, which incorporates correlation and interchannel effects at the single-excitation level, thus corroborates the cutoff extension predicted by Ref.~\cite{Buth2011,Buth2013,Buth2015} at the single atom level.

A schematic of the HHG process yielding to the cutoff extension is depicted in Fig.~\ref{fig:figure_krypton}(b): the $4p$ electron is ionized and accelerated by the IR field. Within half of one IR cycle, the transient valence hole is refilled by the $3d$ electron due to the charge migration dynamics driven by the resonant XUV field. Upon IR field reversal, the valence photoelectron recombines into the $3d$ hole in the cation. The energy difference between the valence and core orbital into  which the $4p$ photoelectron recombines determines the cutoff extension in Fig.~\ref{fig:figure_krypton}(a).

\subsubsection{Extended spectrum in Argon}

We now investigate in more detail the spectral properties of the harmonics lying beyond the regular harmonic cutoff. To keep calculations tractable, we switch to argon as prototype with $3s$ and $3p$ active orbitals resonantly coupled by the XUV field. We use a radial grid extension of $150$ bohr, $L_{max}=80$ and a CAP starting at $130$ bohr. 
All other input parameters are the same as for krypton. Pulse parameters are given in caption of Fig.~\ref{fig:figure1}.

The harmonic yield in argon is shown in Fig.~\ref{fig:figure1}(a). In the absence of the XUV field (red lines)
the energy cutoff corresponds to $\omega^\prime_c\!=\!60.20\,$eV. The XUV+IR configuration
results in the spectrum shown in blue in Fig.~\ref{fig:figure1}(a), with a energy cutoff of $\omega_c=78.86\,$eV corresponding to the harmonic $55$ in Fig.~\ref{fig:figure1}(b). The difference of cutoff energy with and without the XUV is $\omega_c-\omega^\prime_c=\Delta E\!=\!18.66\,$eV, matching the orbital energy difference $\epsilon_{3p}-\epsilon_{3s}$ in the Hartree-Fock approximation. The hole migration sequence giving rise to the cutoff extension is depicted in Fig.~\ref{fig:figure1}(b): it results from the radiative recombination of the $3p$ valence photoelectron into the $3s$ hole in the cation created by the resonant XUV pulse which drives subcycle $3p\rightarrow 3s$ hole transfer within half of the IR period.

\begin{figure}[!t]  
\centering                                                                       
\includegraphics[width=0.98\linewidth]{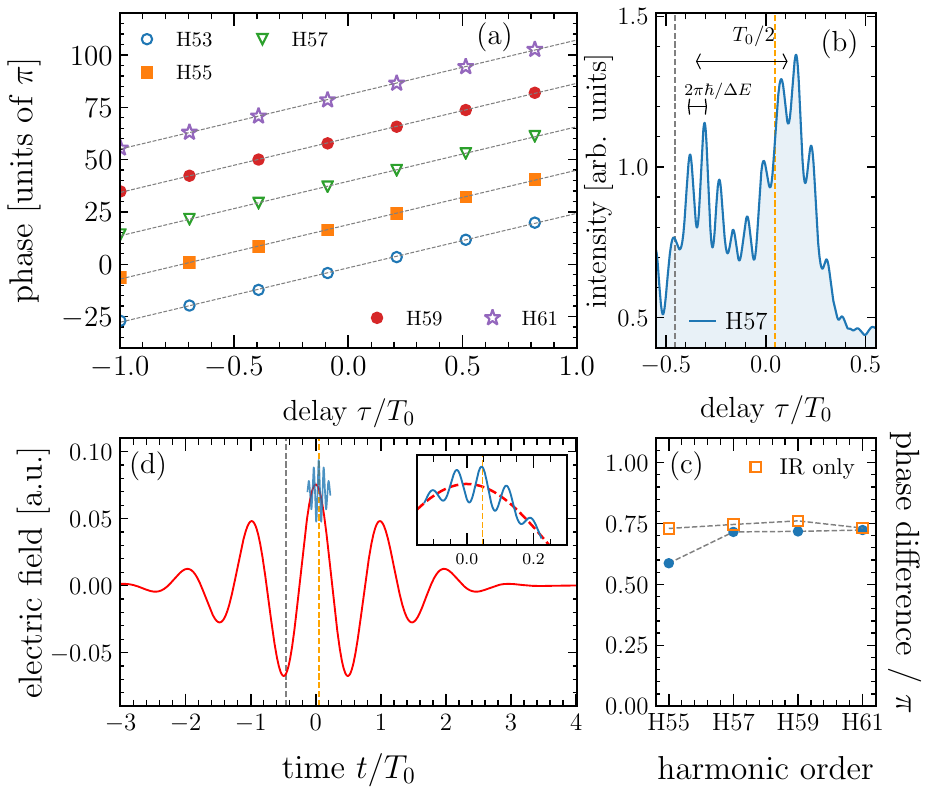}                    
\caption{(a) Phase  and (b) intensity of harmonics $\mathrm{H}_{2j+1}$ vs. delay.
 (c) Spectral phase difference $\Phi_{2j+1}-\Phi_{2j-1}$ vs.  harmonic order for XUV+IR (blue). IR-only (yellow): $\mathrm{H}_{2j+1}$ 
 are defined by the harmonics in the XUV+IR case from which we have subtracted $13\,\omega_0$
 and corresponds to the plateau harmonics of IR-only. (d) IR and XUV field configuration for
 delays $\tau_1=(17^\circ / 360^\circ) T_0$ and $\tau_2=\tau_1-T_0/2$ indicated by the 
 orange ($\tau_1$) and 
 gray ($\tau_2$) vertical lines in (b) and (d).
 IR parameters: 6fs (FWHM), $2\times 10^{14}\,\mathrm{Wcm}^{-2}$ and  $\omega_0\!=\!1.436~$eV. XUV parameters: $0.5~$fs (FWHM), $2\times 10^{13}\,\mathrm{Wcm}^{-2}$ and $\omega_{\mathrm{XUV}}=13\,\omega_0$.
 }
\label{fig:figure2} 
\end{figure}


 The resonant XUV field plays here a similar role to that of the two-body interchannel interactions in the giant enhancement of the HHG spectrum in xenon~\cite{PabsPRL2013}. Specifically, interchannel interactions are mediated by the term  $\hat{H}_1$ in Eq.~\eqref{eqn:com} and, in xenon, it triggers hole dynamics between the $4d$ and $5p$ orbitals. As a result of the recombination of the $5p$ photoelectron into the $4d$ hole in the cation, the HHG yield in xenon is ten-fold enhanced around $100$~eV~\cite{PabsPRL2013}. Figure~\ref{fig:figure1}(c) shows the harmonic spectrum in argon obtained by neglecting (intra) or including (inter) two-electron interchannel interactions. Clearly, interchannel couplings play does not play any role in the enhancement observed in Fig.~\ref{fig:figure1}(a), nor it modifies or suppresses the enhancement predicted by Ref.~\cite{Buth2011,Buth2015} by counteracting the hole dynamics triggered by the XUV field. Within the TDCIS approximation, our results show that the cutoff extension is no spurious and that correlation effects plays an irrelevant role for the cases considered here.  

\begin{figure}[!t]  
\centering                                                                       
\includegraphics[width=0.98\linewidth]{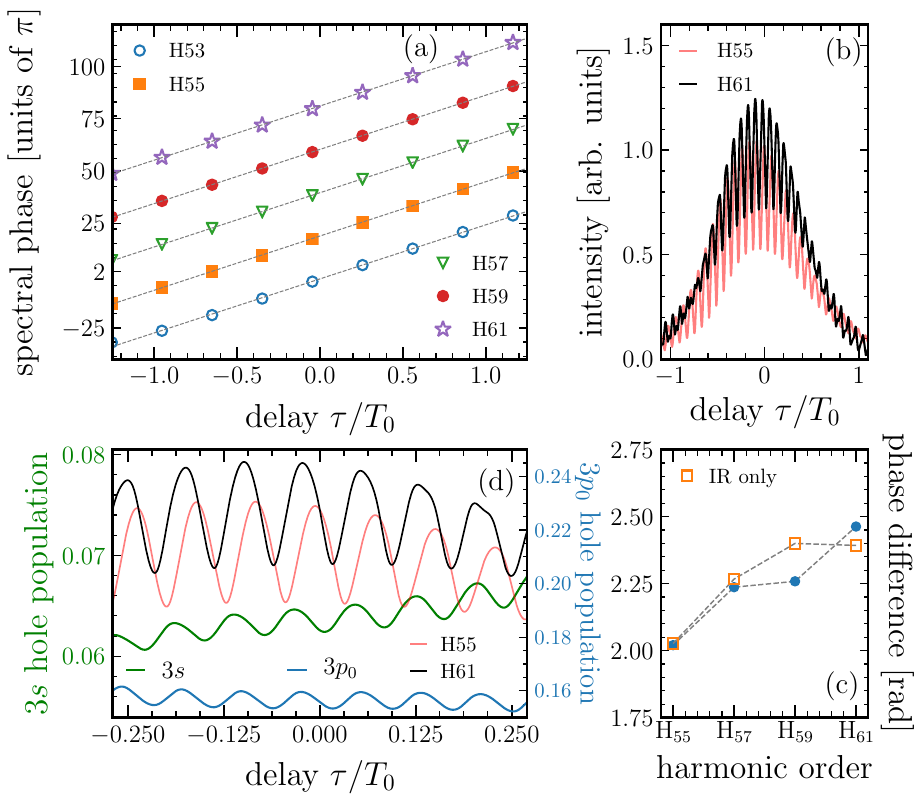}                    
\caption{(a-c) Same as Figs.~\ref{fig:figure2}(a-c) but for 2 fs XUV pulse.
(d) $3s$ and $3p_0$ hole populations at the end of the pulses vs. XUV-IR delay. The intensity profiles of H55 (red) and H61 (black) in (b) are also shown in (d) (not in scale).}
\label{fig:figure2fs} 
\end{figure}

\subsubsection{Spectral Sensitivity on the XUV-IR delay}

Synchronization of the probe XUV and IR pulses as both propagate in the target medium is a sensitive parameter of which depend the mutual coherence of the atomic dipoles generating the macroscopic HHG signal.
Neglecting Gouy-phase and propagation effects, and assuming the refraction indices of the XUV and IR fields equal to unity, the dipole spectrum of two argon atoms located at positions $(0,0,z_i)$ and $(0,0,z_j)$ are related by  $\hat{d}(\omega,z_j)=\exp[i\phi(\omega)]\,\hat{d}(\omega,z_i)$ with $\phi(\omega)=\omega(z_j-z_i)/c$. Since the macroscopic HHG signal consists in the radiation fields generated by each atomic dipole $\hat{d}(\omega,z_k)$ located at $z_k$ that add up coherently at e.g., the sample exit point, it is important for phase matching conditions that the dipole phase distribution change slowly within the sample~\cite{salieres1995}. If $n_{\mathrm{IR}}\!\ne\! n_{\mathrm{XUV}}$, a relative delay $\tau_k = (z_k-z_0)(n_{\mathrm{XUV}}-n_{\mathrm{IR}})/c$ between them would have consequences on the coherence of the emitting dipole distribution as different atoms located at different positions $z_k$, would be perturbed with a different phase, determined by the (position-dependent) delay $\tau_k$. It is thus of importance to analyze how fast the dipole phase change as a function of the XUV-IR delay.

Figure~\ref{fig:figure2}(a) shows the dipole spectral phase sensitivity on the IR-XUV delay $\tau$. A positive (negative) delay indicates that the IR peak intensity is advanced (retarded) with respect to the XUV peak intensity. The spectral phase in Fig.~\ref{fig:figure2}(a) is a linear function of the delay $\tau$ ,
\begin{eqnarray}
\label{eq:linear_fit}
\Phi(\omega;\tau) = \Delta E\,\tau/\hbar + b(\omega)\,,
\end{eqnarray}
where $\Delta E =(\varepsilon_{3p}-\varepsilon_{3s})$ for argon and $b(\omega)$ an offset. The slope $d\Phi(\omega,\tau)/d\tau=\Delta E/\hbar$, which determines the spectral phase sensitivity on the XUV-IR delay, is independent of the harmonic order and solely dictated by the energy difference between the valence and core orbitals coupled by the XUV field, $\Delta E$. The latter is retrieved from the numerical slope $\alpha$ via a fitting procedure $y(\tau)=\alpha\, \tau + b(\omega)$ with $\alpha=28.5\,\mathrm{rads}/\mathrm{fs}$. The phase in Eq.~\ref{eq:linear_fit} is related to the phase accumulated in the ion during ionization and recombination~\cite{smirnova2009strong,smirnova2009high}.

For large $\Delta E$, the spectral phase is a largely sensitive, linear function of the delay. In argon,  a delay $\tau =0.1$~fs suffices for a relative change of $\pi$ in the spectral phase of the enhanced harmonics. Such a sensibility is even more striking in krypton,  for which an
analogous calculation leads to a five-fold steeper slope of $\alpha_{Kr}=137.2\,\mathrm{rad}/\mathrm{fs}$ . In fact, for the excitation scheme in krypton of Fig.~\ref{fig:figure_krypton}(b), $\Delta E=89.82$ eV at the Hartree-Fock level, in contrast to only 18.67 eV for the present excitation scheme in argon. 

\begin{figure}[!t] 
\centering                                                                       
\includegraphics[width=0.99\linewidth]{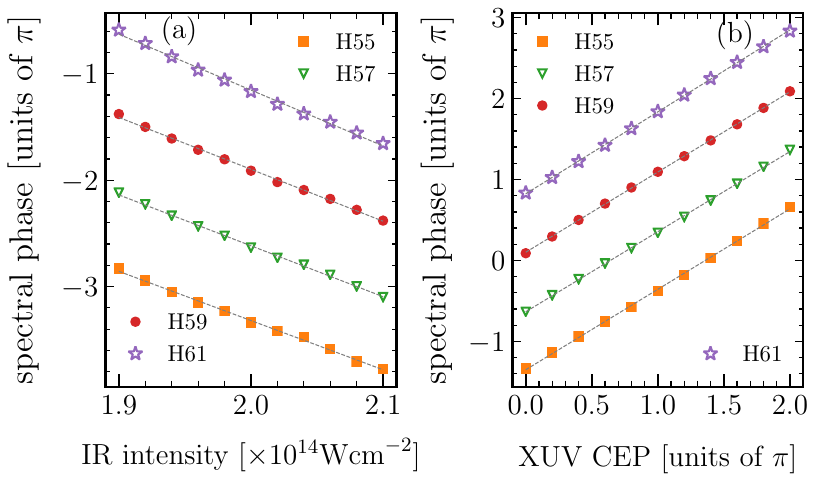}                    
\caption{(a) Spectral phase of harmonics H55 - H61 vs IR intensity. (b) Spectral phase vs XUV CEP phase. Same pulse parameters as in Fig.~\ref{fig:figure2}}.              
\label{fig:figure3} 
\end{figure} 

In contrast to the slope $\alpha\!=\!\Delta E/\hbar$ in Fig.~\ref{fig:figure2}(a), which is independent of the harmonic order,  the offset $b(\omega)$ is not. However, the relative differences between the offsets of two consecutive harmonics are related by the relation,
\begin{eqnarray}
b(\omega_{2n+1})-b(\omega_{2n-1})= C_n\,,
\end{eqnarray}
where $\omega_{2n+1}=(2n+1)\omega_0$ denotes the harmonic order.

In  Fig.~\ref{fig:figure2}(a), the value of $C_n\!\approx\!2.25\,$rad. remains almost constant for all harmonics lying beyond the harmonic cutoff $\omega_c/\omega_0 = 55$, where $\omega_c=78.86\,$eV corresponds precisely to cutoff energy
$\omega_c = |\varepsilon_{3s}| + 3.17 E^2_{\mathrm{IR}}/4\omega^2_0$ with $|\varepsilon_{3s}|$
the ionization potential associated to the $3s$ orbital and $E_{\mathrm{IR}}$ the IR field strength.

The offset difference $b(\omega_{2n+1})-b(\omega_{2n-1})$ defining the spacing between two consecutive parallel lines in Fig.~\ref{fig:figure2}(a) is related to the attochirp~\cite{attochirp1,attochirp2} and reflects the phase difference accumulated during ionization and recombination by the photoelectron giving rise to the enhanced harmonics $\omega_{2n+1}$
and $\omega_{2n-1}$. To confirm this, we show in Fig.~\ref{fig:figure2}(c) the offset difference for the IR+XUV case (solid-blue circles) and the offset difference for the IR only case (empty-orange circles). For the IR-only
case, the harmonics in Fig.~\ref{fig:figure2}(c) correspond  to those lying in the plateau (of the IR-only case) and defined by the XUV+IR enhanced harmonics energy from which we have subtracted $\epsilon_{3p}-\epsilon_{3s}=13\omega_0$. Their values agree within $10\%$, their difference being attributed to the contribution of the XUV field which is absent in the IR-only case. Thus, from a experimental perspective, Fig.~\ref{fig:figure2}(a) showing the slope and offsets of the spectral phase versus delay of the enhanced harmonics represents a real observable giving insights into the ionization and hole dynamics that can be measured using
high-resolution self-referring interferometric HHG spectroscopy techniques~\cite{harrison2023}. 

While the spectral phase in Fig.~\ref{fig:figure2}(a) contains information on the attochirp and hole dynamics through the offsets and slope, respectively, the harmonic
intensity shown in Fig.~\ref{fig:figure2}(b) reveals details of the XUV pulse duration. Figure~\ref{fig:figure2}(b) shows the intensity of H59 as a function of delay for a $0.5\,$fs XUV pulse. The intensity exhibits peaks with period $T_{3s,3p} = 2\pi\hbar/\Delta E= 0.22\,$ fs and modulated by the half the
IR period $T_0/2 \!=\! 1.43\,$ fs. This modulation leads to two distinctive peak envelopes
close to theoretical ionization times $\tau_1\!=\!(17^{\circ}/360^{\circ}) T_0$ and $\tau_2=\tau_1-T_0/2$, indicated by the orange ($\tau_1$) and gray ($\tau_2$) vertical lines. 
\begin{figure}[!t]
\centering                                                                
\includegraphics[width=0.99\linewidth]{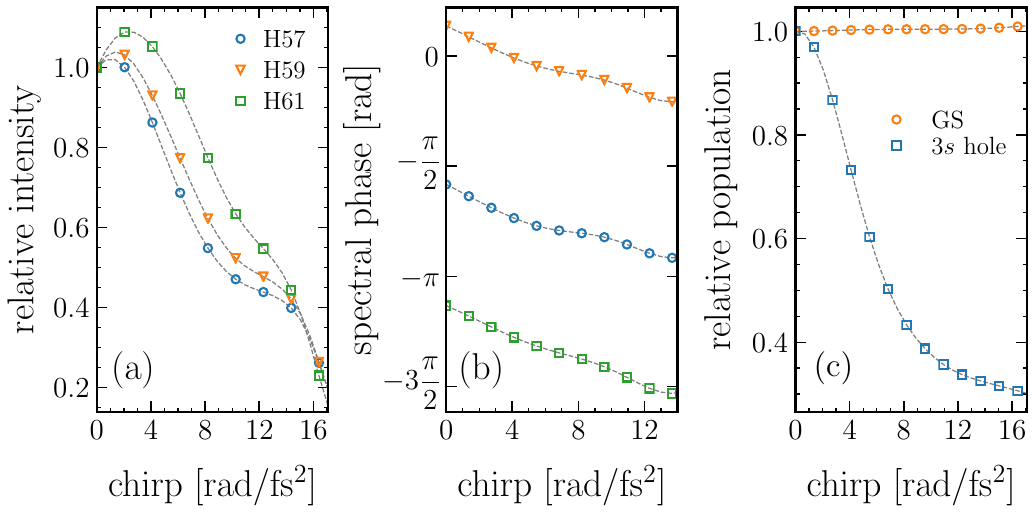}                   
\caption{(a) Relative intensity and (b) spectral phase of harmonics vs. XUV chirp parameter $\beta$. (c) Final-time ground-state (GS) and $3s$ hole population in argon. IR field parameters are the same as in Fig.~\ref{fig:figure1}(a)}
\label{fig:figure4} 
\end{figure}

The difference in the harmonic intensity at $\tau_1$ and $\tau_2$ can be explained by noticing that most of the harmonic signal is generated around $17^\circ$ of the IR peak intensity when $\tau\!=\!0$. Positioning the XUV pulse at $\tau\gg\tau_1$ ensures an efficient $3s\rightarrow 3p$ refilling as ionization has reached its maximum but results in a weaker HHG signal because the IR intensity drops for times $t>\tau_1$. Conversely, positioning the XUV before the IR peak position results in a even weaker HHG signal, because the XUV pulse dies off before the IR reaches its maximum. In this case, ionization of the $3p$ electron is not maximized and the refilling is not efficient. This explains the asymmetric profile around $\tau=0$ in Fig.~\ref{fig:figure2}(b). In both cases, the XUV pulse is not positioned at or slightly after maximum ionization, neither ionization nor hole transfer are optimal. Moreover, at $\tau\!=\!\tau_1$, ionization is enhanced since the IR and XUV peak amplitudes are in phase, as shown in Fig.~\ref{fig:figure2}(c) and inset therein.  The maximum peak position in Fig.~\ref{fig:figure2}(b) is slightly after $\tau_1$. It corresponds to a delay of $\tau_{max}=0.15\,T_0=2\,T_{3s,3p}$. It ensures maximum refilling of the $3p$ hole following ionization and maximum $3s$ hole population at the recombination time.

\begin{figure}[!t]
\centering                                                                
\includegraphics[width=0.90\linewidth]{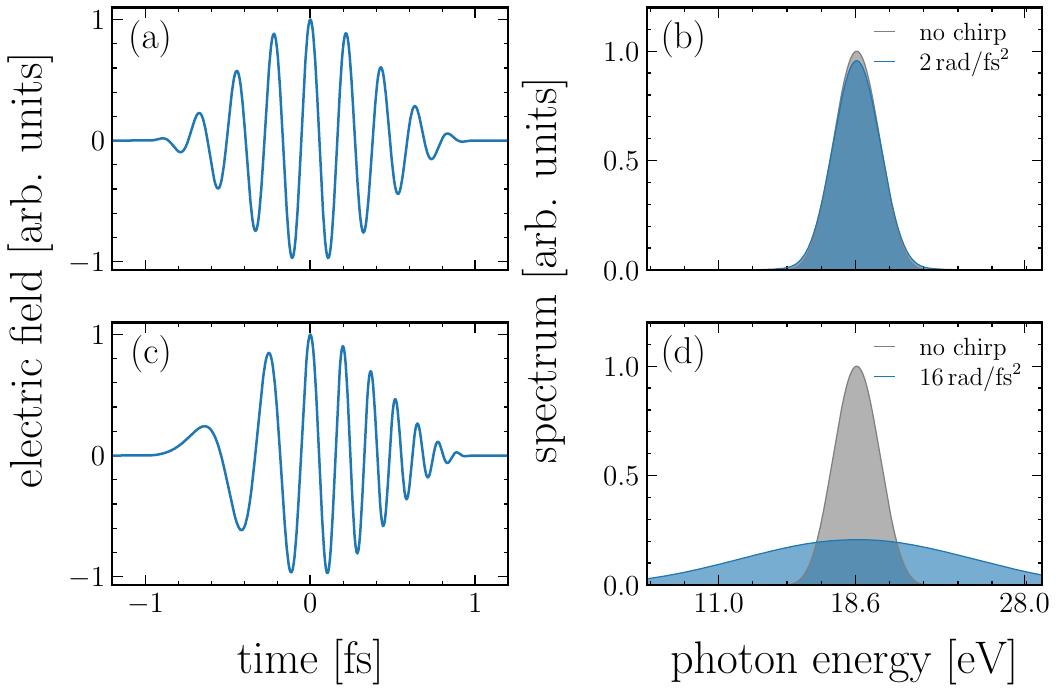}                   
\caption{(a) XUV  pulse with $\beta=2\,\mathrm{rad}/\mathrm{fs}^2$. (b) Frequency distribution of XUV pulse in (a). (c) Same as in (a) but for $\beta=16\,\mathrm{rad}/\mathrm{fs}^2$. (d) Same as in (b) but for $\beta=16\,\mathrm{rad}/\mathrm{fs}^2$.}
\label{fig:figure_chirp} 
\end{figure}

Figures~\ref{fig:figure2fs}(a) and (c) show the same as in Fig.~\ref{fig:figure2}(a), resp. (c) but
for a 2 fs (FWHM) XUV pulse instead of 0.5 fs. The slope $\alpha=\Delta E/\hbar$ for both cases are identical.  The phase difference $\hat{\Phi}(\omega_{2n+1})-\hat{\Phi}(\omega_{2n-1})$ in Fig.~\ref{fig:figure2fs}(a) for the 2 fs XUV differs from that of Fig.~\ref{fig:figure2}(a) but it is still related to the attochirp as shown in Fig.~\ref{fig:figure2fs}(c), where we have applied the same argument as in Fig.~\ref{fig:figure2}(c). Another difference is that the $T_0/2$ modulation of the intensity in Fig.~\ref{fig:figure2}(b) for the 0.5 fs case is no longer present in Fig.~\ref{fig:figure2fs}(b), where only oscillations at $T_{3s,3p}$ show up. Also, the FWHM of the intensity profile in Fig.~\ref{fig:figure2fs}(b) matches that of the IR field. These observations are explained the fact that the 2 fs XUV spans almost one cycle of the IR pulse. Therefore the minimum between $\tau_2$ and $\tau_1$ at $\tau_{min}=-T_0/8$ in Fig.~\ref{fig:figure2}(b) for the 0.5 fs case is lifted up, because the 2 fs XUV does not die off sufficiently fast between within one-eight of $T_0$ to create the double-peak structure as for the 0.5 fs case.  H55 and H61 in Fig.~\ref{fig:figure2fs}(b) exhibit a relative shift of $23\,$as.

Modulation with period $T_{3s,3p}=2\pi\hbar/\Delta E$ in Figs.~\ref{fig:figure2}(b) and~\ref{fig:figure2fs}(b) is related to the state of the cation probed by the recolliding photoelectron~\cite{smirnova2009strong}. Large (small) $3s$ hole population at the successive recollision times results in a high (low) spectral intensity beyond the regular cutoff.
In Fig.~\ref{fig:figure2fs}(d) we show the $3s$ (green) and $3p$ (blue) hole populations evaluated after the XUV and IR pulses are over versus delay. The intensity profiles of H55 (black) and H61 (red)
are also shown for comparison (not in scale). Hole populations and harmonic intensity are both modulated at $2\pi\hbar/\Delta E$
although the harmonic yields are not in phase with the final-time $3s$ hole population nor with the oscillations of the $3s-3p$ degree of coherence (data not shown). This is not unexpected, as the phase of the harmonics contains cumulative contributions of several recombination events.

A linear dependence of the spectral phase is also observed as a function of the IR intensity, cf., Fig.~\ref{fig:figure3}(a), and the XUV carrier-envelope phase, $\varphi_{\mathrm{XUV}}$, c.f., Fig.~\ref{fig:figure3}(b). In the former case, the negative slopes are independent of the harmonic order and  given by $-5.8\,U_p/\omega_0$~\cite{salieres1995} with $U_p=|E_{\mathrm{IR}}|^2/4\omega^2_{IR}$ the ponderomotive energy. The phase variations versus the IR intensity are related to the time needed for the electron to return to the parent atom~\cite{salieres1995}. Also here, the phase spectrum is quite sensitive to IR intensity: a relative IR intensity variation of only $10\%$ leads to a spectral phase variation of $\pi$ for all harmonics in the extended region. Such a rapid variation has significant effects on the macroscopic radiation. It induces phase modulation which leads to spectral broadering of the macroscopic harmonic pulse~\cite{salieres1995}.

Finally, in Fig.~\ref{fig:figure3}(b), the spectral phase dependence on the XUV CEP phase, $\varphi$, is given by $\Phi(\omega;\varphi) = \tilde{\alpha}_{\omega}(\varphi)+\tilde{b}_{\omega}$. The slope can be inferred analytically,
\begin{eqnarray}
    \dfrac{d\Phi(\omega;\varphi)}{d\varphi}=\dfrac{d\Phi(\omega;\tau)}{d\tau}\times \dfrac{d\tau}{d\varphi}=\left( \dfrac{\Delta E}{\hbar}\right) \left(\dfrac{\hbar}{\Delta E}\right) = 1,\quad
\end{eqnarray}
where $d\Phi(\omega;\tau)/d\tau\!=\!\Delta E/\hbar$, previously found in Eq.~\eqref{eq:linear_fit}, and $\tau=\phi/\omega=\phi/(\Delta E/\hbar)$ with $\omega=\omega_{\mathrm{XUV}}=\Delta E/\hbar$. Numerical evaluation of the slope in Fig.~\ref{fig:figure3}(b) confirms that the spectral phase follows linearly the XUV CEP with a calculated numerical slope  $\Tilde{\alpha}_{\omega}(\varphi)=1$ for  all enhanced harmonics. The offsets in Fig.~\ref{fig:figure3}  increase linearly as a function of the harmonic order.

\subsubsection{Chirped XUV Pulses}

We consider here XUV pulses with an intrinsic chirp defined by a
time-varying optical phase of the form $\Phi_{\mathrm{XUV}}(t)=\omega_{\mathrm{XUV}} t+\beta t^2$. 
Reported values of up to $\beta=76\,\mathrm{rad}/\mathrm{fs}^2$ obtained by quadratic fitting of the spectral phase between harmonics 11 and 19 has been reported in argon at $10^{14}\mathrm{Wcm}^{-2}$~\cite{intrinsic_chirp} and 
intrinsic attochirps of $45\,\mathrm{as/eV}$ ($\beta=36.5\,\mathrm{rad}/\mathrm{fs}^2$) at
$0.5\,\mu$m and $21.5\,\mathrm{as/eV}$ ($\beta=70.5\,\mathrm{rad}/\mathrm{fs}^2$) at $2\,\mu$m~\cite{attochirp_8as_eV}.
In what follows, the XUV envelope is fixed, described by the Gaussian function in Eq.~\ref{eqn:plw}, and independent of $\beta$.

Figure~\ref{fig:figure4} investigates
the impact of chirped XUV pulses on the intensity and spectral phase of the extended harmonics in argon
with panel (a) showing the harmonic intensity as a function of $\beta>0$.
The intensity values are normalized with respect to the unchirped case ($\beta=0$).

Interestingly, a chirp rate of $\beta\approx 2\,\mathrm{rad/fs}^2$ results in a $10\%$ intensity increase 
for H61, and a much lower increase ($\approx 1\%$) for H57 and H59. The temporal profile of the corresponding XUV pulse can be seen in Fig.~\ref{fig:figure_chirp}(a). Such a increase is rather surprising because the XUV spectral intensity at exactly the theoretical $3s\rightarrow 3p$ transition energy $\Delta E$ is 4\% weaker for the chirped pulse compared to the unchirped case as shown in Fig.~\ref{fig:figure_chirp}(b). This is because for a chirped pulse the frequency distribution spreads out while keeping the integrated spectral density constant.  Consequently, the intensity at $\Delta E$ decreases as $\beta$ increases.  This weakens the coupling between both hole states.  The increase with respect to the unchirped case
can be attributed to an optimal timing for refilling of the transient $3p$ hole 
following ionization as the instantaneous XUV frequency $(d/dt)\,\Phi_{\mathrm{XUV}}(t)$ evolves in time before matching the resonant
transition energy in the presence of the dressing IR field.

\begin{figure}[!t] 
\centering                                                                
\includegraphics[width=0.99\linewidth]{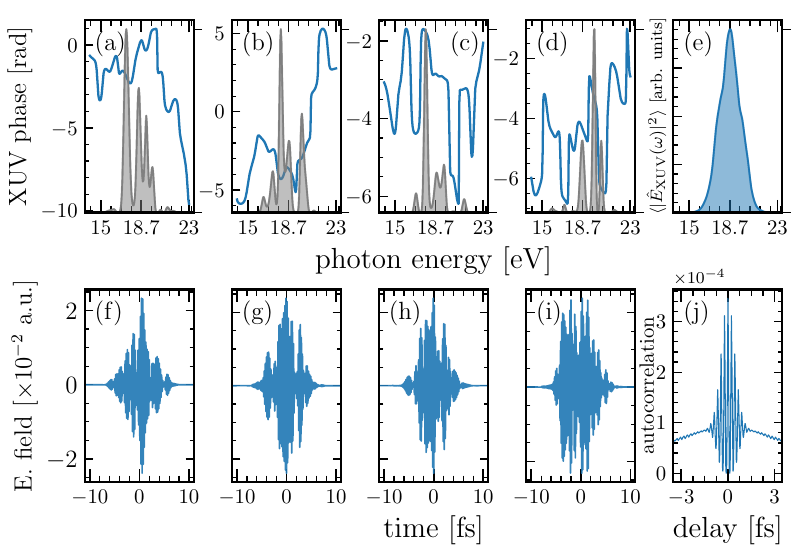}                    
\caption{(a-d) Shot-to-shot spectral intensities (filled gray) and spectral phases (blue lines) of partially-coherent XUV fields. (e) Shot-averaged spectral intensity. (f-i) Time profile of the XUV pulses corresponding to panels (a-d). (j) Shot-averaged autocorrelation function.}     
\label{fig:figure6v2} 
\end{figure} 

Ultimately, as $\beta$ increases, the harmonic intensities drop significantly, reaching a eight-fold suppression for
 $\beta>16\,\mathrm{rads/fs}^{2}$, c.f.,~\ref{fig:figure4}(a). The  corresponding spectral
phase evolution, shown in Fig.~\ref{fig:figure4}(b), exhibit a much slow, quasi-linear decrease. Fig.~\ref{fig:figure_chirp}(c) shows the time profile of the chirped pulse with $\beta=16\,\mathrm{rad}/\mathrm{fs}^2$ and Fig.~\ref{fig:figure_chirp}(d) its spectral components (blue). For comparison, the spectral density of the unchirped pulse is also displayed in Fig.~\ref{fig:figure_chirp}(d) (gray). 

The intensity at the resonance energy $\Delta E$ decreases dramatically for $\beta=16\,\mathrm{rad}/\mathrm{fs}^2$, which explains the suppression of the harmonic yield shown in Fig.~\ref{fig:figure4}(a).
To confirm this, the final-time ground-state depletion and the $3s$ hole population 
is further analyzed in Fig.~\ref{fig:figure4}(c).  
The final-time $3s$ hole population is maximum for the unchirped case and exhibits a rapid and monotonic decrease for large chirp rates while the ground-state population remains almost unaltered. This is explained by the suppression of $3s\rightarrow 3p$ channel that refills the $3p$ hole created upon weakly ionization of $3p$ valence electrons for large values of $\beta$. Thus, efficiency to
drive the required hole dynamics by the XUV pulse is significantly suppressed as indicated by the $3s$ hole population shown in Fig.~\ref{fig:figure4}(c).  With a broader spectral density distribution and subsequent weaker intensity at the resonance energy due to the chirp, the refilling mechanism is suppressed. 

\begin{figure}[!t]                                
\centering                                                                
\includegraphics[width=0.99\linewidth]{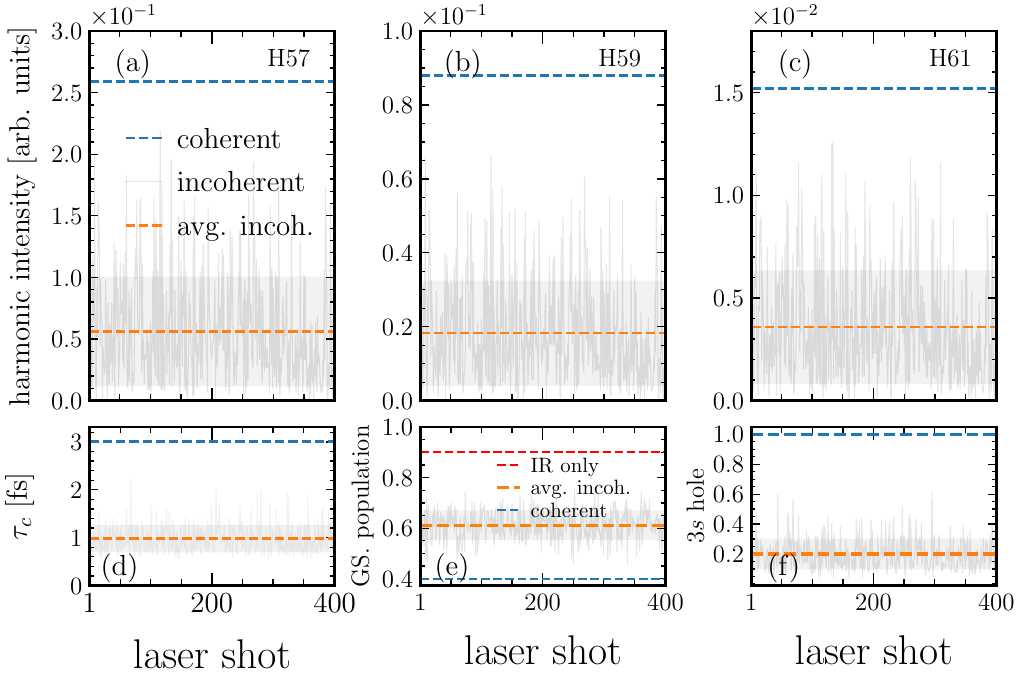}                    
\caption{(a) Shot-to-shot intensity of H57, (b) H59 and (c) H61 obtained with coherent (blue) and partially-coherent (gray) XUV fields. Intensity is normalized with respect to cutoff H45 of IR-only. Filled gray curves indicates standard deviation to the mean value (orange). (d) Shot-to-shot coherence time $\tau_c$ (gray) and averaged coherence time $\langle\tau_c\rangle$ (orange). (e) Final-time ground-state population. (f) Final-time 3s hole population. }  
\label{fig:figure5} 
\end{figure}

\subsection{Partially coherent XUV pulses}

In this section, we analyze ability of partially-coherent XUV pulses, defined according to Eq.~\eqref{eqn:pcm}, to efficiently drive the hole dynamics required to enhance the harmonic yield in the extended region.

Figure~\ref{fig:figure6v2} displays the coherence properties of a few XUV pulses generated following the PCM approach~\cite{Pfeifer2010}, with
panels (a) to (d) showing the spectral density (filled gray) and spectral phase (blue lines). Even though the shot-to-shot spectral density distribution may vary significantly, the shot-averaged spectral density (averaged over $400$ shots), shown in panel (e), tends to a Gaussian-like function peaked at the desired $3s-3p$ energy. 
Panels (f) to (i) show
the electric field profile given by Eq.\eqref{eqn:pcm} associated to panels (a) to (d), respectively. Figure~\ref{fig:figure6v2}(j) displays the shot-averaged 
autocorrelation function averaged over all the electric fields in the statistical sample.
The width of the sharp center peak in Fig.~\ref{fig:figure6v2}(j) is determined by the bandwidth of the shot-average spectral density and determines the coherence time $\tau_c$~\cite{Pfeifer2010}.

Figure~\ref{fig:figure5} compares the microscopic harmonic yield of the extended harmonics obtained
with coherent and partially-coherent XUV pulses. The coherent XUV field corresponds to a bandwidth-limited
pulse in accordance with Eq.\eqref{eqn:plw} with the same pulse parameters as in Fig.~\ref{fig:figure1}(a)
but with a FWHM of $4$ fs. Partially-coherent XUV pulses are generated according to Eq.~\eqref{eqn:pcm}. In either case, the IR field is assumed to be coherent with same pulse parameters as in Fig.~\ref{fig:figure1}(a).

 Solid-gray lines in Fig.~\ref{fig:figure5} indicate the HHG yield obtained with partially-coherent XUV pulses for different laser shots. Each shot configuration is defined by one XUV pulse generated with the PCM approach while keeping the IR field unaltered. The shot-averaged HHG yield (avg. incoh in Fig.~\ref{fig:figure5}) has been averaged over 400 shots and its average value is highlighted by the dashed-orange lines in Fig.~\ref{fig:figure5}. 
 For comparison, the dashed-blue lines corresponds to the intensity obtained with the coherent (bandwidth-limited) $4$-fs XUV pulse. Harmonic yields are normalized with respect to the regular energy cutoff (IR only) at H45.
 
 For the pulse parameters adopted here, we have found a five-fold suppression of the intensity of H57 for partially-coherent pulses compared to the coherent counterpart. As similar ratio is found for H59 shown in Fig.~\ref{fig:figure5}(b), and a ratio of $3.8$ for H61, whose intensity is already
two orders of magnitude weaker than the IR-only harmonic cutoff for the coherent XUV pulse case. In all panels, filled-gray lines indicates the standard deviation from the mean values shown in orange.

Figure~\ref{fig:figure5}(d) displays the XUV coherence times $\tau_c$, c.f., Eq.~\eqref{eqn:tau_c}, as a function of the laser shot (gray-lines). The shot-averaged coherence time, $\langle\tau_c\rangle$, shown in orange, is approximately $1\,$fs, whereas the coherence time for the coherent XUV pulse is $3\,$fs. It is worth mentioning that the harmonic intensity is not a monotonically decreasing function of $\tau_c$ for every shot configuration but compared to the coherent XUV case, it leads to a statistically-averaged weaker HHG signal for a statistically-averaged smaller coherence time $\langle\tau_c\rangle$.

Finally,  Figs.~\ref{fig:figure5}(e) and (f) show the shot-averaged ground-state and
$3s$ hole population, respectively, obtained with the coherent and partially-coherent XUV pulses. Similarly to the discussion for chirped pulses (which incidentally have also limited temporal coherence), the efficiency
to drive the $3s\rightarrow 3p$ transition is, on average, five-fold inferior for partially-coherent pulses compared to the bandwidth-limited counterpart, c.f., Fig.~\ref{fig:figure5}(f). This inability to effectively drive the required hole population transfer results in a decrease of the final-time $3s$ hole population, from $\approx 1$ for the coherent case to only 0.2 for the partially coherent case, as shown in Fig.~\ref{fig:figure5}(f) in blue and orange, respectively. Within the PCM model, the five-fold suppression of the harmonics compared to the coherent case can thus by explained at the microscopic level by the off-resonant character of the shot-to-shot spectral density distribution of partially-coherent fields even though the averaged XUV spectrum is peaked at the right target transition energy.

\section{Macroscopic Effects}
\subsection{Theory}

In this section we analyze the influence of propagation effects on the harmonics lying beyond the regular energy cutoff (IR only). Specifically, we analyze the effects of modification of the incident IR and resonant XUV field during propagation in a macroscopic medium on the macroscopic harmonic signal. Propagation of the incident field is dictated by the Maxwell equation~\cite{LHuillier1991},
\begin{eqnarray}
\label{eq:maxwell1}
    \left(\nabla^{2}-\dfrac{\partial^2}{\partial t^2}\right)\boldsymbol{E}(\boldsymbol{r},t) = \dfrac{4\pi}{c^2}\dfrac{\partial^2}{\partial t^2}\boldsymbol{P}(\boldsymbol{r},t)
\end{eqnarray}
where $\boldsymbol{E}(\boldsymbol{r},t)$ denotes the propagating electric field which depends
on the polarization
$\boldsymbol{P}(\boldsymbol{r},t)$, which in turn depends on $\boldsymbol{E}(\boldsymbol{r},t)$.
In order to provide a estimate of the influence of the propagation effects, we resort to integrate Eq.~\eqref{eq:maxwell1} in one dimension only and assume that the transverse compoenent of the incident XUV+IR field are not modified during propagation. With this assumption, the inhomogeneous wave equation in frequency domain can be recast in the following form,
\begin{eqnarray}
\label{eq:maxwell2}
    \left(\dfrac{\partial^2}{\partial z^2}+k^2(\omega)\right) \hat{E}(z,\omega) = -4\pi\dfrac{\omega^2}{c^2}N_V\, \hat{d}_{nl}(z,\omega)\,,
\end{eqnarray}
with $\omega$ the circular frequency, $c$ the speed of light in vacuum, $N_V$ the density of atoms per unit volume, and  $k(\omega)=\omega n(\omega)/c$ with $n(\omega)=\sqrt{(1+4\pi\chi^{(1)})}$ the refractive index where $\chi^{(1)}(\omega)=N_V \alpha(\omega)$ is the linear susceptibility with $\alpha(\omega)$ the atomic polarizability. In deriving Eq.~\eqref{eq:maxwell2}, we have separated
the polarization into its linear and nonlinear functional components, $P(z,\omega)=P_L(z,\omega)+P_{NL}(z,\omega)$~\cite{LHuillier1991} with $P_{L}(z,\omega)=N_V\,d_l(z,\omega)$ and $P_{NL}(z,\omega)=N_V\, d_{nl}(z,\omega)$ with $d_l(z,\omega)=\alpha(\omega)\,E(z,\omega)$ resp. $d_{nl}(z,\omega)$ the linear and nonlinear components of the microscopic (atomic) dipole moment at frequency $\omega$ of the atom located at the sample point $z$.

In general, $d(z,\omega)=d_{l}(z,\omega)+d_{nl}(z,\omega)$ depends on all amplitudes $E(\omega_1,z), E(\omega_2,z),\, ...\, E(\omega_N,z)$. Here, we neglect the influence of the generated high-order harmonics owing their weak intensities and assume that the response of the atom at position $z$
depends on field amplitudes $E(z,\omega_1)$ and $E(z,\omega_X)$ only, where $\omega_1$ and $\omega_X$ denote the frequencies of the incident IR and resonant XUV, i.e.,
\begin{eqnarray}
\label{eq:ddd}
    d(z,\omega) = d\left(z,\omega, E(\omega_1,z), E(\omega_X,z)\right)\,,
\end{eqnarray}
which, to avoid cumbersome notations, we write simply as $d(z,\omega)\equiv d(z,\omega,E(z,\omega_1),E(z,\omega_X))$, with $E(z,\omega)$ denoting the total field amplitude at frequency $\omega$ and at position $z$ in the sample.

In a XUV+IR configuration, there are two mechanisms that are absent in the regular HHG (IR only):  lost of temporal coherence of the propagating pulse due to polarization-induced difference in the propagation speed and absorption of the incident XUV field.
If the IR and XUV fields propagates at the same speed (same refractive index), the dipole at the entrance point $z_0$ and that at $z$ are related by $d(z,\omega)\!=\!d(z_0,\omega)\,\exp[i\omega n(\omega_1)(z-z_0)/c]$. In contrast, different IR and XUV refractive indices, $n(\omega_1)$, resp. $n(\omega_X)$ results in a position-dependent delay $\tau(z)=(n(\omega_1)-n(\omega_X))(z-z_0)/c$  between both pulses, which according to Fig.~\ref{fig:figure2}, introduces an additional
phase $\Phi(z,\omega)=\Delta E/\hbar\times  \tau(z)$ to the dipole $d(z,\omega)$, this is
\begin{eqnarray}
\label{eq:phaseg}
    \Phi(z,\omega) = \dfrac{\Delta E}{\hbar} \times \dfrac{\omega}{c}\left [n(\omega_1)-n(\omega_{XUV})\right](z-z_0),
\end{eqnarray}
which modifies the regular phase matching conditions (IR only)~\cite{LHuillier1991}. The phase~\eqref{eq:phaseg} originates from 
the relative propagation speed between the IR and resonant XUV field and depends on the energy difference $\Delta E$, which we recall, corresponds to the energy difference between the valence and inner shell orbitals (e.g., $3s$ and $3p$ in argon) in the Hartree-Fock picture. The influence of this phase can be controlled by adjusting the sample length or, alternatively, the gas density, as $n(\omega_1)$ and $n(\omega_{XUV})$ depends on $N_V$ as $n(\omega)=\sqrt{1+4\pi N_V \alpha(\omega)}$.

\begin{figure}[!t]                                                                      
\centering                                                                       
\includegraphics[width=0.99\linewidth]{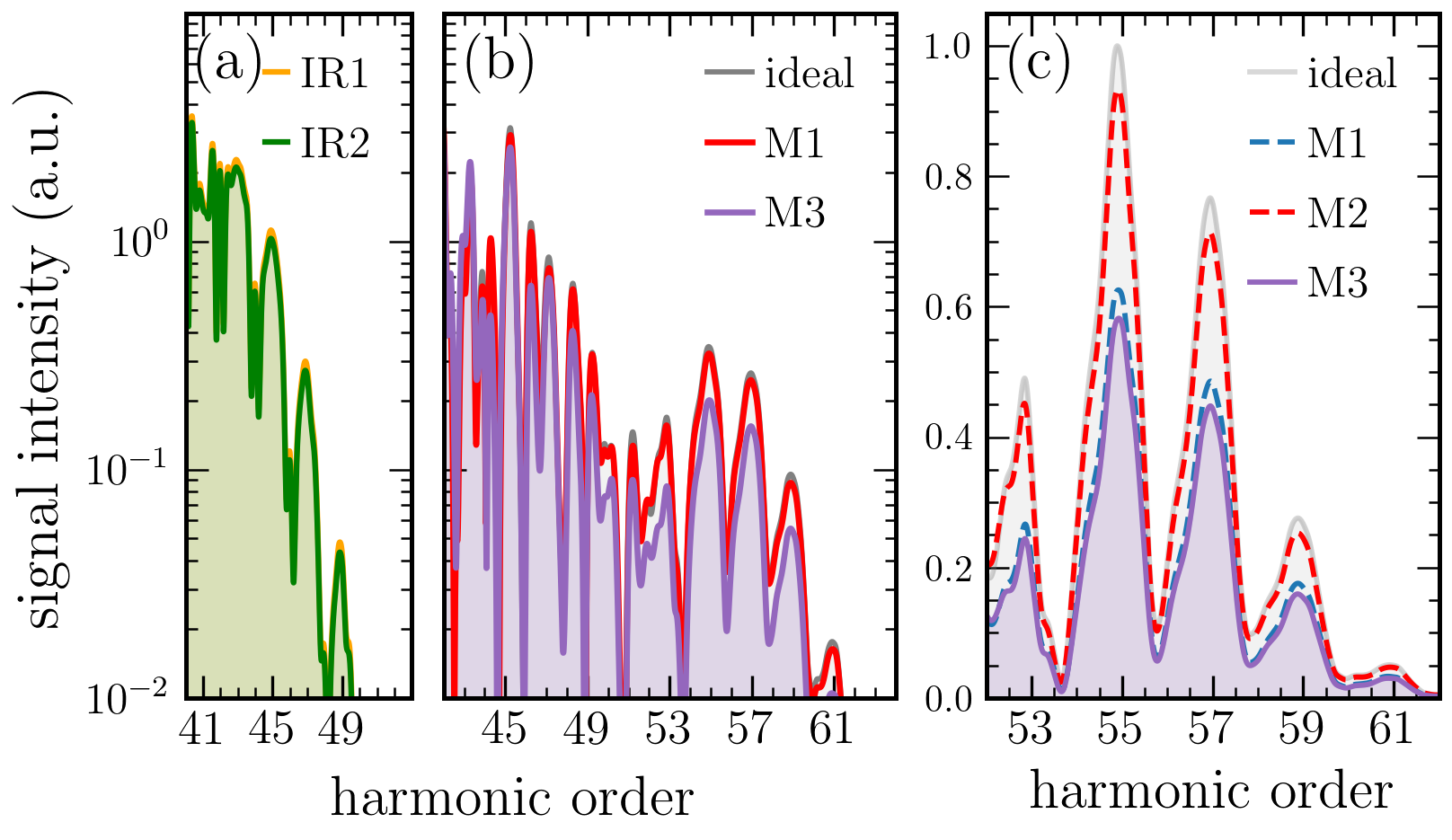}                    
\caption{(a) Macroscopic HHG spectrum in Argon for a sample length of $L\!=\!0.5\,$mm at $20\,$Torr in the absence of the resonance XUV field. Two cases are shown: perfect phase matching (IR1) and IR with dispersion (IR2). (b) HHG spectrum in the presence of the XUV pulse for different propagation models, c.f., text. (c) Same as panel (b) but in linear scale showing in more detail the contributions of the different propagation effects to the signal suppression.}
\label{fig:macroscopic_1} 
\end{figure}

The other mechanism that affects the macroscopic signal beyond the regular cutoff region 
is the absorption of the resonant XUV frequency $\omega_{XUV}$. It induces excitation of the valence electrons as well as $3p\rightarrow 3s$ hole transfer if the $3p$ orbital is vacant.  The microscopic HHG yield at the extended harmonic orders is a linear function of the XUV intensity (data not shown). Absorption at the incident XUV frequency corresponds to the imaginary part of $n(\omega_{XUV})$ via $\alpha(\omega_{XUV})$.

To account for the effect of phase mismatch (retardation) between the incident XUV and IR fields through $\Phi(z,\omega)$ in Eq.~\eqref{eq:phaseg} and absorption of the XUV, we perform one-dimensional model calculations that accounts
for these effects explicitly. In fact, under the assumption~\eqref{eq:ddd}, Eq.~\eqref{eq:maxwell2} can be solved analytically,
\begin{eqnarray}
\label{eq:maxwell3}
    E(z,\omega_{q}) &\!=\!&2i\pi\dfrac{\omega}{cn_q} N_V\, e^{ik_q z}\int^{z}_{z_0} e^{-ik_q}\, d_{nl}(z^\prime,\omega_q)\, dz^\prime\,,\quad\quad
\end{eqnarray}
with $k_q=n_q\omega_q/c$ where $n_q$ denotes the refractive index at the extended harmonic frequency $\omega_q$.  To obtain the harmonic field amplitude $E(z,\omega_q)$ in Eq.~\eqref{eq:maxwell3}, we generate
independent TDCIS calculations for a set of XUV and IR pulses describing the fields  $E(z^\prime,t)$ at each
position $z^\prime$ in the sample according to Eqs.~\eqref{eqn:plw} and~\eqref{eqn:plww}. From each independent TDCIS calculation we extract the resulting dipole $d_{nl}(z^\prime,\omega_q)$ appearing in Eq.~\eqref{eq:maxwell3}. The refractive index of the XUV ($n(\omega_{XUV})$) and that of the generated harmonics ($n_q$) are set to unity while that of the IR ($n(\omega_1)$) are obtained from Ref.~\cite{rahman1990dynamic}. The XUV field intensity at each position $z^\prime$ is generated by accounting for the XUV absorption coefficients from Ref.~\cite{nist_xray}. We use a density of $1000$ TDCIS calculations per milimeter of sample length. Retardation of the IR with respect to the XUV and attenuation of the XUV due to absorption depend both on the gas pressure. 

\begin{figure}[!t]                                                                      
\centering                                                                       
\includegraphics[width=1.0\linewidth]{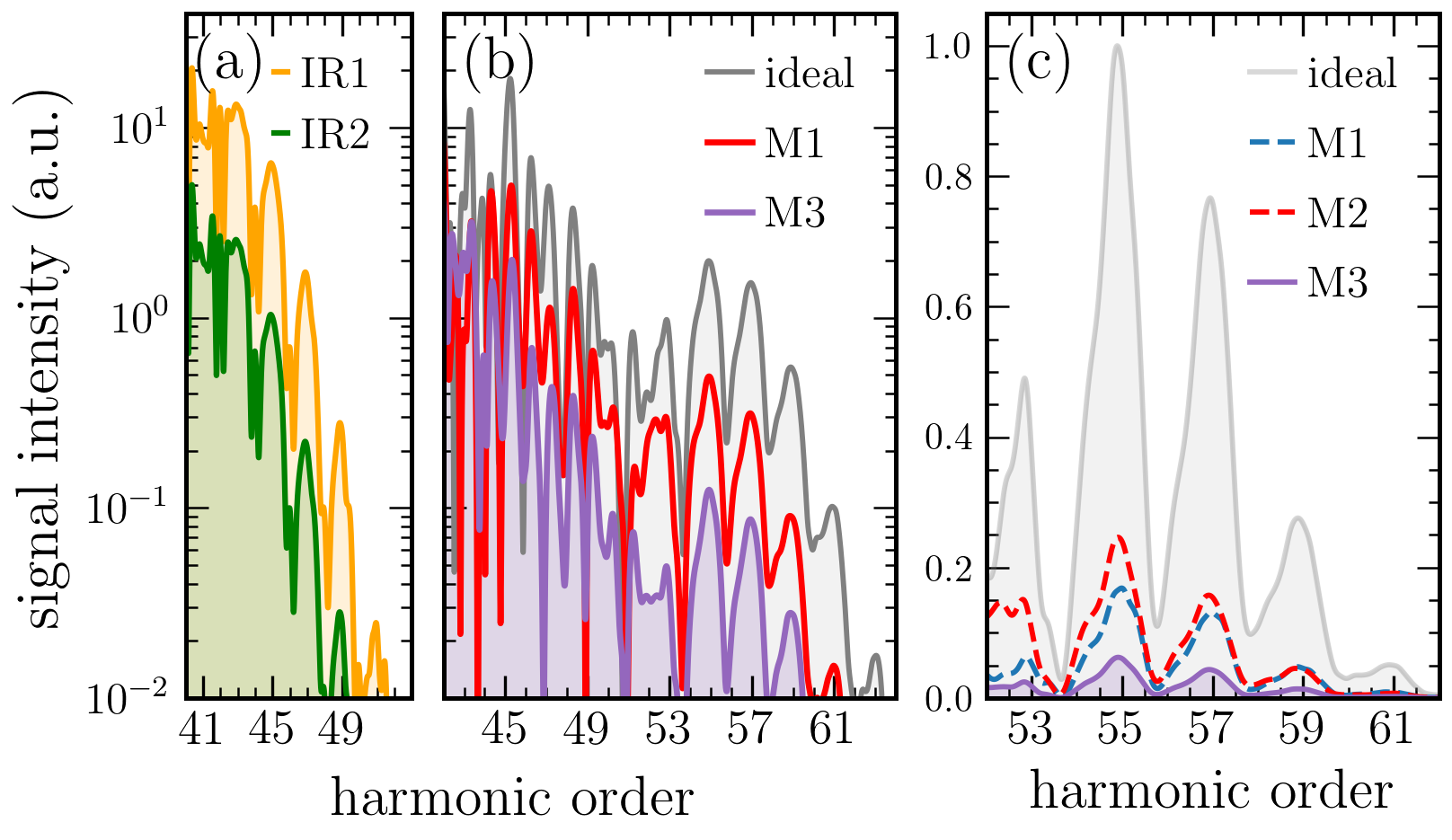}                    
\caption{Same as Fig.~\ref{fig:macroscopic_1} but for a sample length and gas pressure configuration of $L\!=\! 1.0\,$mm and $P\!=\! 40\,$Torr. As in Fig.~\ref{fig:macroscopic_1}, the incident XUV and IR field parameters at the entrance point $z_0$ of the sample correspond to those of Fig.~\ref{fig:figure2}.
}        
\label{fig:macroscopic_2} 
\end{figure}

\subsection{Numerical Results}

To analyze the effects of XUV absorption and IR retardation on the HHG spectrum, we calculate the harmonic amplitude~\eqref{eq:maxwell3} at the sample exit point according to three different propagation models. In model M1, we take into account absorption of the resonant XUV but neglect retardation of the IR field. Conversely, in model M2, we account for retardation of the IR field by means of an refractive index but neglect absorption of the XUV. Note that M2 accounts for the regular phase mismatch $\Delta k_q = \left(n(\omega_q)-n(\omega_1)\right)\omega_q/c$ between the IR of frequency $\omega_1$ and generated harmonics $\omega_q=q\omega_1$~\cite{LHuillier1991} as well as the additional phase $\hat{\Phi}(z,\omega)$  of Eq.~\eqref{eq:phaseg} shown in Figs.~\ref{fig:figure2}(a) and~\ref{fig:figure2fs}(a). In addition to the phase $\hat{\Phi}(z,\omega)$, the delay between the XUV and IR results in the modulation of the atomic dipole intensity shown in Figs.~\ref{fig:figure2}(b) and~\ref{fig:figure2fs}(b). Finally, we account for both effects (XUV absorption and IR retardation) in model M3. For comparison, we also show results for the "ideal" case, representing the case with neither XUV absorption nor IR retardation.

Figure~\ref{fig:macroscopic_1}(a) shows the harmonic spectrum in Argon with (IR2) and without (IR1) phase mismatch between the generated harmonics and the incident IR field in the absence of the resonant XUV pulse (IR only). Sample length and gas pressure are $L\!=\!0.5$ mm and $P\!=\!20$ Torr, respectively. For IR1 (yellow) we use $n(\omega_q)\!=\!n(\omega_1)\!=\! 1.0$. For IR2 (green), we use $n(\omega_q)\!=\!1.0$ and $n(\omega_1)-1\!=\!6.4905\!\times \!10^{-6}$. Phase mismatch results in a suppression of about $9\%$ relative to the case of perfect phase matching,  with a ratio of $\mathcal{R}_0=1.09$ at the harmonic cutoff H45. 

When the XUV is switched on, in Fig~\ref{fig:macroscopic_1}(b), the intensity at H45 is reduced by a factor of $\mathcal{R}_0=1.08$ when comparing the "ideal" case against model M2 (unmodified XUV, retarded IR). A ratio of $\mathcal{R}_H=1.04$ is found for the harmonics beyond the regular cutoff. When accounting for XUV absorption (M1) in Fig.~\ref{fig:macroscopic_1}(b), the ratio becomes $\mathcal{R}_0=1.2$ at H45 and $\mathcal{R}_H=1.75$ in the extended region. The contribution of XUV absorption (M1) and IR retardation (M2) on the overall signal suppression can be seen in Fig.~\ref{fig:macroscopic_1}(c), showing the spectral intensity in the extended region for the different models. For this sample length and gas pressure configuration, absorption of the resonant XUV is the driving mechanism for suppression of the HHG yield beyond the regular cutoff energy.

We now consider a sample length and gas pressure configuration corresponding to 
$L\!=\!1.0\,$mm and $P\!=\!40\,$Torr ($n(\omega_1)-1\!=\!1.2981\times 10^{-5}$). 
Figure~\ref{fig:macroscopic_2}(a) shows the effect of phase mismatch near the regular 
harmonic cutoff H45 in the absence of the resonant XUV field (IR only). 
In this case,  phase mismatch between the IR and generated harmonics (IR2) 
results in a suppression of the harmonic yield by a factor of $\mathcal{R}_0\!=\!7.8$ at H45 relative to the case of perfect phase matching (IR1).
When the resonant XUV is switched on, in Fig.~\ref{fig:macroscopic_2}(b), 
retardation effects (M2) results in a suppression factor of $\mathcal{R}_0\!=\!3.6$ at H45 relative to the "ideal" case.  In the extended region, IR retardation results in a suppression of about $\mathcal{R}_H\!=\!5$.  When XUV absorption and IR retardation effects are simultaneously included (M3), the harmonic yield in extended region is suppressed by a factor $\mathcal{R}_H=17.76$ relative to the "ideal" case.
If we resort to normalize the intensities for the XUV+IR case relative to the yield 
at the harmonic cutoff H45 in the IR-only case (IR2), as performed  in
Figs.~\ref{fig:macroscopic_1} and~\ref{fig:macroscopic_2}, the signal suppression 
in the extended region 
relative to the IR-only scenario then corresponds to  $\mathcal{R}_H=1.75$ for $L=0.5\,$mm  and $P=20\,$Torr 
and $\mathcal{R}_H=17.76$ for $L=1.0\,$mm and $P=40\,$Torr. 

Finally, Fig.~\ref{fig:macroscopic_2}(c) shows
the contributions of XUV absorption (M1), IR-retardation (M2) and their combined effects (M3) on the signal suppression. We find that both effects 
 have commensurate contributions in the extended region, resulting in a suppression 
factor of $17.75$ relative 
to the signal intensity at H45 in the IR-only case. It is also to note that, in addition to the suppression due to propagation effects considered in this section, 
further suppression is expected for partially-coherent XUV pulses, with an additional suppression factor of 5 as discussed in Sec.~\ref{subsec:partially_coherent}.

\section{Conclusions} 
\label{sec:conclusions}
In conclusion, we have analyzed the coherence properties of high-order harmonic radiation 
enhanced by XUV-driven hole population transfer by incorporating multi-electron interchannel and correlation effects at the configuration-interaction single level.
We have focused our analysis on the sensibility of
the harmonics extending beyond the regular cutoff energy in standard high-order harmonic generation.
In particular, we have shown that the dipole spectral phase is particularly sensitive to the XUV-IR time delay. Such a time delay can naturally occur if both the XUV and IR propagates at different speed in the sample due to modification of their refractive indices. This feature modifies the standard phase matching conditions. We have shown that the dependence of the microscopic spectral phase on the IR-XUV delay is determined by the energy difference $\Delta E$ between the inner and valence orbitals resonantly coupled by the XUV field. While such a sensibility is expected to be even more striking 
for the $3d\rightarrow 4p$ transition
in krypton than in argon, we have shown using a one-dimensional model that the suppression of the extended harmonics at the macroscopic level due to phase mismatch between the propagating XUV and IR fields play a minor for low gas pressures and short sample lengths. We have also shown that absorption of the incident resonant XUV results in a non-negligible suppression of the macroscopic harmonic signal at the extended photon energies even for perfectly coherent XUV pulses. Such a effect can be further exacerbated for partially-coherent XUV pulses, where we obtained a five-fold suppression compared to the coherent case. We have shown that such a signal suppression can be interpreted at the single-atom level by the inability for chirped and XUV pulses with poor statistical temporal coherence
to efficiently drive the required hole dynamics. 

Our work demonstrates that propagation effects and the temporal coherence of the resonant XUV pulse are important for the description of XUV-assisted high-order harmonic generation as experimentally probed in Ref.~\cite{Tross2022}. In this context, a complete three-dimensional description for the field propagation and inclusion of the spatial distribution of Laguerre-Gaussian beams in addition to the partial temporal coherence of the resonant XUV needs to be accounted for fully explain the experimental observations of Ref.~\cite{Tross2022}.

\subsection*{Acknowledgements} 

Financial support from the U.S.Department of Energy (DOE), Office of Science,
Basic Energy Sciences (BES) under Award Number DE-SC0023192 is gratefully acknowledged.


\bibliography{bibmanuscript}

\end{document}